\documentclass[a4paper,11pt]{article}
\pdfoutput=1 

\usepackage{jheppub} 

\usepackage[T1]{fontenc} 
\usepackage{slashed,color,graphicx}
\usepackage{float}
\usepackage{subfig}
\usepackage{amsmath}
\usepackage{amssymb}
\usepackage{bbold}
\usepackage[dvipsnames]{xcolor}
\usepackage{centernot}

\usepackage{cancel}

\newcommand{\la}{\lambda}

\newcommand{\wt}{\widetilde}
\newcommand{\nl}{\nonumber\\}
\newcommand{\bea}{\begin{eqnarray}}
\newcommand{\eea}{\end{eqnarray}}

\begin{document}

\title{\boldmath Implications of CDF $W$-mass and $(g-2)_\mu$ on  $U(1)_{L_\mu-L_\tau}$ model}


\author{Seungwon Baek,}
\emailAdd{sbaek1560@gmail.com}
\affiliation{Department of Physics, Korea University, \\
Anam-ro 145, Sungbuk-gu, Seoul 02841, Korea}

\abstract{
    We study the implications of the recent anomalies in the $W$-boson mass and the anomalous magnetic moment of the muon
on $U(1)_{L_\mu-L_\tau}$ model. We show that the introduction of vector-like { leptons} which mix with muon can solve both anomalies.
Contrary to the conventional wisdom the electroweak scale $Z'$-boson is allowed without conflict with the trident neutrino production experiments.
}

 \maketitle 
 
 \section{Introduction}
\label{sec:intro}
Although the standard model (SM) has been very successful for decades, there are some hints indicating the existence of new physics (NP).
The measurement of the anomalous magnetic moment of the muon, $a_\mu=(g-2)_\mu/2$, from the Muon $g-2$ Collaboration  at Fermilab~\cite{Muong-2:2021ojo}, when combined with the value reported by the E821 Collaboration at the Brookhaven National Laboratory~\cite{Muong-2:2006rrc}, deviates from the SM prediction~\cite{Aoyama:2020ynm}\footnote{We refer the reader to the references in \cite{Aoyama:2020ynm} for more theoretical contributions.} by $4.2\sigma$,
\bea
\Delta a_\mu = a_\mu^{\rm exp} - a_\mu^{\rm SM} = (25.1 \pm 5.9) \times 10^{-10},
\label{eq:Delta_amu}
\eea
although more precise nonperturbative QCD calculation is required in the theory side. The deviation may be explained by a NP which couples to the muon.

The $U(1)_{L_\mu-L_\tau}$ extension of the SM gauge group~\cite{He:1990pn,He:1991qd} has received much attention~\cite{Baek:2008nz,Baek:2015mna,Baek:2015fea,Baek:2017sew} as, among other things, the contribution of the new gauge boson, $Z'$, can solve the discrepancy~\cite{Baek:2001kca}. However the experimental data for the cross section of the neutrino trident production, the scattering of the muon neutrino off heavy 
nuclei producing a $\mu^+\mu^-$ pair, or the process  $\nu_\mu N \to \nu_\mu N \mu^+ \mu^-$ in the SM, turn out to be very stringent in constraining the parameter space of $U(1)_{L_\mu-L_\tau}$ models.  The CHARM-II Collaboration~\cite{CHARM-II:1990dvf} and the CCFR Collaboration~\cite{CCFR:1991lpl} detected the events with the cross sections
\bea
\frac{\sigma^{\rm CHARM-II}}{\sigma^{\rm SM}}=1.58 \pm 0.57 \quad\text{and}\quad
\frac{\sigma^{\rm CCFR}}{\sigma^{\rm SM}}=0.82 \pm 0.28,
\label{eq:trident_exp}
\eea
which are in good agreement with the SM expectations. 
Applying the 2$\sigma$ upper bound from the CCFR result, the authors of~\cite{Altmannshofer:2014pba} found that only a small region with $m_{Z'} \sim {\cal O}(10^{1\pm 1})$ MeV and the gauge coupling $\sim {\cal O}(10^{-4})$ can explain the $\Delta a_\mu$ in 
the  minimal model of $U(1)_{L_\mu-L_\tau}$, excluding the electroweak scale gauge boson of $U(1)_{L_\mu-L_\tau}$ among other things.
The model can be extended, for example, to incorporate the neutrino masses and mixings to explain the neutrino oscillation data~\cite{Baek:2015mna,Baek:2015fea}.

Recently the CDF Collaboration has announced the measurement of the W-boson mass~\cite{CDF:2022hxs}
\bea
m_W = 80,433.5 \pm 9.4 \, {\rm MeV},
\label{eq:CDF_MW}
\eea
which is in tension with the SM prediction $m_W^{\rm SM} = 80,357 \pm 6$ MeV from the precision electroweak data by 7$\sigma$.
This may also call for NP if it is confirmed by the future experiments~\cite{Fan:2022dck,Lu:2022bgw,Athron:2022qpo,Yuan:2022cpw,Strumia:2022qkt,Yang:2022gvz,deBlas:2022hdk,Du:2022pbp,Tang:2022pxh,Cacciapaglia:2022xih,Blennow:2022yfm,
Arias-Aragon:2022ats,
Sakurai:2022hwh,Fan:2022yly,Liu:2022jdq,Lee:2022nqz,Cheng:2022jyi,Bagnaschi:2022whn,Paul:2022dds,Bahl:2022xzi,Asadi:2022xiy,DiLuzio:2022xns,Athron:2022isz,Gu:2022htv,Heckman:2022the,Babu:2022pdn,Heo:2022dey,Du:2022brr,Cheung:2022zsb,Crivellin:2022fdf,Endo:2022kiw,Biekotter:2022abc,Balkin:2022glu,Ahn:2022xeq,Han:2022juu,Zheng:2022irz,Kawamura:2022uft,Ghoshal:2022vzo,Perez:2022uil,Mondal:2022xdy,Zhang:2022nnh,Borah:2022obi,Chowdhury:2022moc,Arcadi:2022dmt,Cirigliano:2022qdm,Carpenter:2022oyg,Popov:2022ldh,Ghorbani:2022vtv,Du:2022fqv,Bhaskar:2022vgk}.

In this paper we show that a minimal extension of the $U(1)_{L_\mu-L_\tau}$ with vector-like leptons allows electroweak scale $Z'$ to explain the $\Delta a_\mu$ and solves the CDF W-mass anomaly at the same time. In Section~\ref{sec:model} we introduce the model. In Section~\ref{sec:gm2} we show that there is large region of parameter space which can solve $\Delta a_\mu$
while evading the constraints from the neutrino trident production and the LHC experiments. In Section~\ref{sec:ST} we calculate the oblique parameters $S$ and $T$ and show that the excess in the $W$-mass with respect to the SM prediction can be accommodated.

\section{The Model}
\label{sec:model}

We introduce vector-like leptons (VLLs) which transform like {$L \sim (1, 2, -1/2, 2)$} and {$E \sim (1, 1, -1, 2)$} in the gauge 
group $SU(3)_C \times SU(2)_L \times U(1)_Y \times U(1)_{L_\mu-L_\tau}$. 
The $U(1)_{L_\mu-L_\tau}$ symmetry is
broken spontaneously when the new scalar field {$\Phi \sim (1,1,0,1)$} develops the vacuum expectation value (VEV).
By definition the second (the third) generation 
left-handed doublet $\ell_\mu$ ($\ell_\tau$) and right-handed singlet $\mu_R (\tau_R)$ leptons of the SM also have $U(1)_{L_\mu-L_\tau}$ charge 
{$+1 (-1)$}.  
 All the rest SM fields are neutral under the new $U(1)$ group. Then the bare masses of the VLLs and the new Yukawa interactions are allowed
\bea
{\cal L} \supset && -m_L \overline{L} L - m_E \overline{E} E   \nl
&& -y_L  \overline{L} P_R E H -y_E \overline{L} P_L E H  +H.c. \nl
&&-\la_L  \overline{L} P_L \ell_\mu \Phi -\la_E \overline{\mu} P_L E \Phi^* +H.c.,
\eea
where $H \sim (1,2,1/2,0)$ is the SM Higgs doublet scalar. We assume the new Yukawa couplings are real parameters.
We can decompose the VLL doublets as $L_{L(R)}=(\wt{N}, \wt{E})^T_{L(R)} $. 

After the breaking of $SU(2)_L \times U(1)_Y$ and $U(1)_{L_\mu-L_\tau}$ symmetries by $\langle H^0 \rangle = v/\sqrt{2}$ and
$\langle \Phi \rangle = v_\Phi/\sqrt{2}$, respectively,
the fermion mass matrix in the basis $(\mu, \wt{E}, E)$ is in the form,
\bea
 {\cal L}_{\rm mass} 
&=& -(\overline{\mu} \; \; \overline{\wt{E}} \; \; \overline{E}) 
\begin{pmatrix}
m_\mu & 0 & {\la_E v_\Phi \over \sqrt{2}}\\
{\la_L v_\Phi \over \sqrt{2}} & m_L & {y_E v \over \sqrt{2}} \\
0 & {y_L v \over \sqrt{2}} & m_E
\end{pmatrix}
P_L
\begin{pmatrix}
\mu \\
\wt{E} \\
E
\end{pmatrix} + H.c.
\label{eq:mass_matrix}
\eea
Denoting the above $3 \times 3$ charged lepton mass matrix as ${\cal M}$,
we can diagonalize it by biunitary transformation,
\bea
V_R^\dagger {\cal M} V_L = \hat{\cal M},
\label{eq:M_diag}
\eea
where  $V_{L,R}$ are real orthogonal matrices, and  $\hat{\cal M}$ is diagonal with positive eigenvalues.
We denote the mass eigenstates as $( E_1, E_2, E_3)$ {and the corresponding} masses { as}
$(M_1, M_2, M_3)$ with { $E_1$ and $M_1$ identified as the mass eigenstate of the muon $\mu'$ and its mass, $m_{\mu'}=0.105658$ GeV, measured in experiments, respectively}.
To guarantee we always get an eigenvalue $m_{\mu'}$ we use the procedure presented in Appendix~\ref{sec:mu_mass}.

\section{The Muon $(g-2)$}
\label{sec:gm2}

\begin{figure}[htb]
      \begin{center}
	\includegraphics[width=0.6\linewidth]{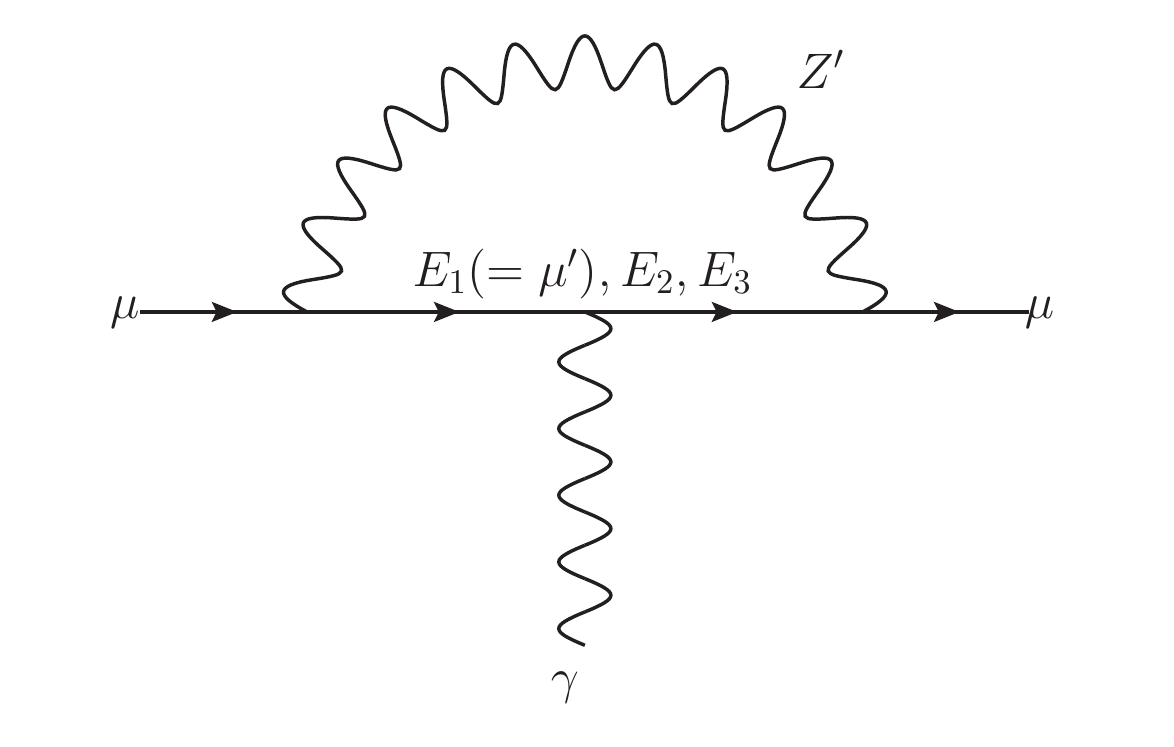}
     \end{center}
	\caption{Feynman diagram for the muon $(g-2)$.}
\label{feynman_gm2}
\end{figure}

Without introducing the VLLs, the minimal model predicts the contribution of the $U(1)_{L_\mu-L_\tau}$ gauge boson, $Z'$,  to the muon $(g-2)$ to be~\cite{Baek:2001kca} 
\bea
a_\mu^{Z'} = \frac{   g_X^2 m_\mu^2}{4 \pi^2} \int_0^1 dx \frac{x^2(1-x)}{m_{Z'}^2 (1-x) + m_\mu^2 x}.
\label{eq:gm2_minimal}
\eea
In the presence of the VLLs which mix with $\mu$ the new contribution can be enhanced due to the chirality flip in the
VLL line with enhancement factor $M_{2(3)}/m_\mu$, which can be huge~\cite{Crivellin:2019mvj}.  
In Fig.~\ref{feynman_gm2} we show the Feynman diagram for the $Z'$ contributions to $(g-2)_\mu$.
There are also the corresponding diagrams with $Z'$ replaced by $Z$. But the $Z$-mediated $\mu$-mixing with $E_{2,3}$ are suppressed compared to
that of $Z'$-mediated $\mu$-mixing with $E_{2,3}$. But we included all the $Z^{(')}$ contributions in our numerical calculation in the following sections.

{
Including the VLL contributions, the NP contribution to the muon $(g-2)$ is given by~\cite{Leveille:1977rc}
\bea
a_\mu^{Z^{(\prime)}} = \frac{m_\mu^2}{4 \pi^2}
\sum_{j} \int_0^1 dx \frac{ ({C^V_{1j}(Z^{(\prime)})})^2  f(x,M_j,m_{Z^{(\prime)}}) + ({C^A_{1j}(Z^{(\prime)})})^2 f(x,-M_j,m_{Z^{(\prime)}}) 
}
{x M_j^2 + (1-x) m_{Z^{(\prime)}}^2 -x(1-x) m_\mu^2}
\label{eq:a_mu_zp}
\eea
where $m_{\mu'} (\equiv M_1)$ is the physical muon mass, $C^{V(A)}_{1j}=(C^R_{1j}\pm C^L_{1j})/2$ ,and $j=1,2,3 (2, 3)$ for the $Z' (Z)$ contribution. 
The couplings $C^{L(R)}_{1j}(Z^{(\prime)})$ and the loop function $f$ are listed in the Appendix~\ref{app:Candf}. The NP contribution, $a_\mu^{\rm NP}$, is the sum of the $Z'$ and $Z$ contributions
\begin{align}
a_\mu^{\rm NP} = a_\mu^{Z'} + a_\mu^{Z}.
\label{eq:a_mu_NP}
\end{align}
When $M_{2(3)} \gg m_\mu$}, in a good approximation we obtain~\cite{Leveille:1977rc} 
\bea
a_\mu^{{ Z'}} &\simeq& \frac{ g_X^2 m_\mu}{16 \pi^2 m_{Z'}^2} \sum_{i=2,3}  M_i V_{L_{1i}} V_{R_{1i}},
\eea

Before presenting the results we consider a couple of relevant constraints. The effective Yukawa interaction between the Higgs boson and the muon pair in our model is given by
\bea
{\cal L} \supset - m_\mu {V_L}_{11} {V_R}_{11} h \overline{\mu^\prime} \mu^\prime,
\eea
where $m_\mu$ is the mass parameter in (\ref{eq:mass_matrix}), which is not necessarily equal to the physical mass $m_{\mu'}$ due to mixing.
Recently the CMS Collaboration has found the ``evidence for Higgs boson decay to a pair of muons''~\cite{CMS:2020xwi}. This measurement gives a constraint
\begin{align}
0.8 < \left| m_\mu {V_L}_{11} {V_R}_{11} \over m_{\mu'} \right| <1.6.
\label{eq:Higgs-to-muon}
\end{align}

The $Z$-boson interaction with the muon pair has been measured precisely at per mille level, whose effective vector-  and axial-vector-couplings are  fitted to be~\cite{Workman:2022ynf}
\begin{align}
\bar{g}_V^\mu = -0.0366 \pm 0.0023, \quad \bar{g}_A^\mu = -0.4994 \pm 0.0005,
\label{eq:Z-mu-mu_exp}
\end{align}
at 1$\sigma$ level.
In our model the coupling deviates from the SM prediction at tree level, which we can read from (\ref{eq:V-m-m}),
\begin{align}
\Delta \bar{g}_V^\mu &={1 \over 2} \left(\left|{V_L}_{31}\right|^2 - \left|{V_R}_{21}\right|^2 \right), \nl
\Delta \bar{g}_A^\mu &={1 \over 2} \left(\left|{V_L}_{31}\right|^2 + \left|{V_R}_{21}\right|^2 \right).
\end{align}
We impose 2$\sigma$ allowed range of (\ref{eq:Z-mu-mu_exp}).

There are lower bounds on the masses of VLLs from the colliders. The most stringent constraint comes from the analyses of the multilepton final states at the LHC.
In the Ref.~\cite{CMS:2017wua}, the CMS Collaboration searched for signals of a type-III seesaw mechanism in events with three or more electrons or muons with the data sample of 35.9 fb$^{-1}$ collected at $\sqrt{s} = 13$ TeV. In the type-III seesaw seesaw mechanism $SU(2)_L$ triplet VLLs with $Y=0$ mix with the SM leptons to generate the
neutrino masses. The charge and neutral VLLs, $\Sigma^\pm, \Sigma^0$, can be pair-produced, $p p \to \Sigma^+ \Sigma^-, \Sigma^\pm \Sigma^0$. 
From their subsequent decays the masses of the VLLs below 850 GeV are excluded~\cite{CMS:2017wua}.
In our model we don't have the final states including electrons, and the pair-production cross sections are also expected to be suppressed due to the mixing between
the doublet and singlet VLLs.
Given the absence of a dedicated study, as far as we are aware of, on the collider constraint on our model we adopt the above 850 GeV as
a conservative lower bound for the VLL masses. 

\begin{figure}[]
	\centerline{\includegraphics[width=0.6\linewidth]{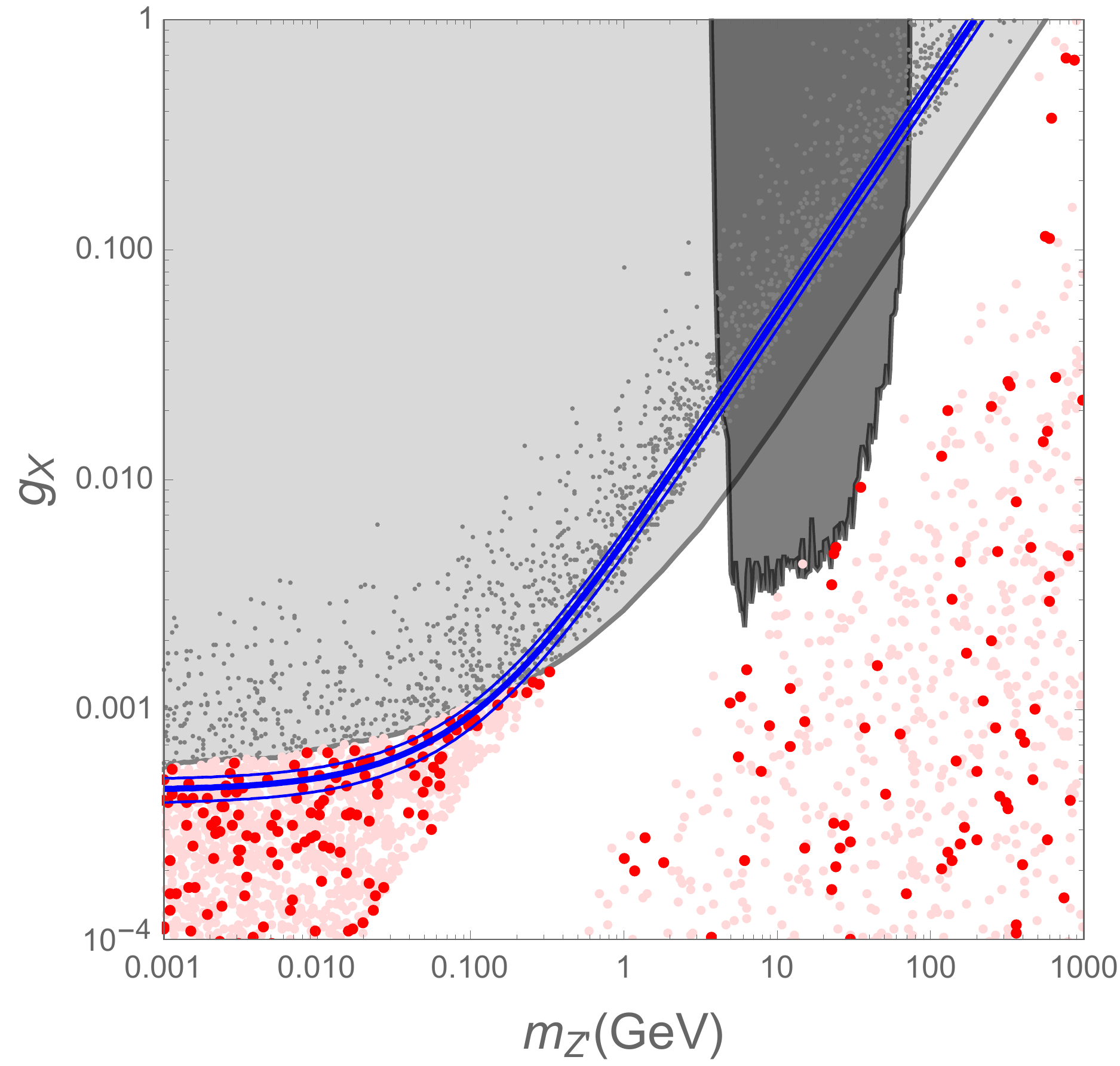}}
	\hspace{0.5cm}
	\caption{Scatter plot in the $(m_{Z'},  g_X)$-plane.
	The light gray region is excluded  by the CCFR measurement of the neutrino trident production cross section at 95\% C.L.. The dark gray region is disfavored by the observation of the SM $Z$ boson decay into four leptons by the CMS Collaboration at the LHC. We show the predictions of the minimal model without the VLLs
	leptons  with blue contours. The thick and thin contour lines give the central and $\pm 1 \sigma$ values of Eq.~(\ref{eq:Delta_amu}), respectively. 
	Tiny gray points are excluded by the trident and/or CMS constraints.
 Light red points are allowed by $(g-2)_\mu$ but disfavoured by the CDF $M_W$.
 Red points are allowed by both the $\Delta a_\mu$ and the CDF $M_W$. See the text for the constraints when generating the points.}
\label{fig:trident}
\end{figure}

Fig.~\ref{fig:trident} shows the prediction of $(g-2)_\mu$ in the $(m_{Z'}, g_X)$-plane along with constraints.
For this plot we scanned in the region
\begin{align}
g_X &\in (10^{-5},1), \nl
m_{Z'} &\in (10^{-3},10^3) \; (\rm GeV), \nl
m_{L,E}&\in (850,5000)\; (\rm GeV), \nl
\lambda_{L,E}, y_{L,E}& \in (-\sqrt{4 \pi}, \sqrt{4 \pi}).
\label{eq:scan}
\end{align}

We collected 5000 points satisfying  1$\sigma$ range of $(g-2)_\mu$,
\begin{align}
19.2 \times 10^{-10} < a_\mu^{\rm NP} <  31.0 \times 10^{-10},
\end{align}
as well as (\ref{eq:Higgs-to-muon}), and (\ref{eq:Z-mu-mu_exp}). 
They are divided into three categories: (A) those excluded by the trident and/or CMS constraints (represented by tiny gray points in Fig.~\ref{fig:trident}), 
(B) those allowed by $(g-2)_\mu$ but disfavoured by the CDF $M_W$ whose precise meaning will be given in the next Section (represented by light red points in Fig.~\ref{fig:trident}), and (C)
those allowed both by $(g-2)_\mu$ and by the CDF $M_W$ (represented by red points in Fig.~\ref{fig:trident}).

The light gray region is excluded  at 95\% C.L. by the CCFR measurement of the neutrino trident production cross section. The dark gray region is disfavored by the observation of the SM $Z$ boson decay into four leptons at the CMS Collaboration at the LHC~\cite{CMS:2012bw}. 
We also show the predictions of the minimal model without the VLLs,  obtained from Eq.~(\ref{eq:gm2_minimal}), with blue contours. 
The thick and thin contour lines give the central and $\pm 1 \sigma$ values of Eq.~(\ref{eq:Delta_amu}), respectively.  They are consistent with the plot in Ref.~\cite{Altmannshofer:2014pba}.  We can see $m_{Z'} \gtrsim 100$ MeV region of the minimal model is disfavored by the trident experiments.

In our model we can see many points (light red and red points) even with electroweak scale $m_{Z'}$ can accommodate $\Delta a_\mu$ from VLL contributions while satisfying the neutrino trident experiments. There are some points (red points) which can satisfy both $\Delta a_\mu$ and the CDF $M_W$.  
Although the process of the neutrino trident production has contributions from the VLLs, their contributions are always subleading due to the fact that there is no enhancement proportional to $M_2,M_3$ contrary to the $(g-2)_\mu$. And the contribution from the virtual muon in the diagram is dominant, and we neglected the VLL contribution in the calculation of the trident cross section.

We can see a peculiar hollow corridor along the diagonal in the middle of the scattered points. It can be roughly understood as follows. For a given $m_{Z'} \gtrsim 1$ GeV, the points above the hollow corridor are obtained mainly by the muon contribution with small ${V_{L(R)}}_{12}$ and/or ${V_{L(R)}}_{13}$, making the VLL contributions subdominant.
For the VLL contributions to be sizable,  it turns out that $v_{\Phi} (= m_{Z'}/g_X) \gtrsim 1 $ TeV is required to generate the sizable mixing with the muon and the VLLs. 
This explains the hollow corridor.

\section{CDF II $W$-mass}
\label{sec:ST}

By including the CDF II $W$-mass measurement in (\ref{eq:CDF_MW}) in the fitting, the authors of \cite{Strumia:2022qkt,Bagnaschi:2022whn} found that the excess can be explained by enhancing especially the oblique $T$-parameter of the electroweak precision tests.
We calculated the $S,T$-parameters in our model. The NP contribution can be obtained from the Lagrangian terms
\bea
{\cal L} \supset && -{g \over \sqrt{2}} W_\mu^+ \overline{\wt{N}} \gamma^\mu \wt{E} + H.c. \nl
&&-\left( -{g' \over 2} B_\mu + {g \over 2} W^3_\mu\right) \overline{\wt{N}} \gamma^\mu \wt{N} 
-\left( -{g' \over 2} B_\mu - {g \over 2} W^3_\mu\right) \overline{\wt{E}} \gamma^\mu \wt{E} 
-\left( -g' B_\mu \right) \overline{E} \gamma^\mu E. 
\eea
Using $P_{L(R)} \wt{E} = V_{L_{2i}} P_{L(R)} E_i$, $P_{L(R)} E = V_{L_{3i}} P_{L(R)} E_i$, we get~\cite{Cynolter:2008ea}
\begin{align}
T =& \frac{1}{4\pi s_W^2 c_W^2 m_Z^2}  \Bigg[ 
2 \sum_{i=2,3} \Big\{(V_{L_{2i}}-V_{R_{2i}})^2 \wt{\Pi}_{LL}(M_0,M_i,0) +  V_{L_{2i}}V_{R_{2i}} \wt{\Pi}_{VV}(M_0,M_i,0)\Big\} \nl
&-\sum_{i,j=2,3} \Big\{ (g_L^{ij}(W^3)-g_R^{ij}(W^3))^2 \wt{\Pi}_{LL}(M_i,M_j,0)
+g_L^{ij}(W^3)g_R^{ij}(W^3) \wt{\Pi}_{VV}(M_i,M_j,0)\Big\} 
\Bigg], 
\end{align}
and
\begin{align}
S =& {1 \over \pi} \Bigg[ \wt{\Pi}'_{VV}(M_0,M_0,0) \nl
&- \sum_{i,j=2,3} \Big\{ (g_L^{ij}(B)-g_L^{ij}(B))(g_L^{ij}(W^3)-g_L^{ij}(W^3)\wt{\Pi}'_{LL}(M_i, M_j,0) \nl
&+ {1 \over 2} (g_L^{ij}(B)g_R^{ij}(W^3)+g_L^{ij}(W^3)g_R^{ij}(B))\wt{\Pi}'_{VV}(M_i,M_j,0)\Big\}
\Bigg],
\end{align}
where $g_{L(R)}^{ij}(W^3) = V_{{L(R)}_{2i}} V_{{L(R)}_{2j}}$ and 
$g_{L(R)}^{ij}(B) = V_{{L(R)}_{2i}} V_{{L(R)}_{2j}} +2 V_{{L(R)}_{3i}} V_{{L(R)}_{3j}}$.
The explicit expressions for the vacuum polarization amplitudes and their derivatives, $\wt{\Pi}$ and $\wt{\Pi}'$, are listed in the Appendix~\ref{sec:VP}.
In the above calculation we neglected the contributions from the mixing between $\mu'$ and $\wt{E}, E$ because they are suppressed either by the small muon
mass compared to $M_2, M_3$ or by the small mixing angles. We have checked the divergences are cancelled and the final results are finite, which validates our calculation.

\begin{figure}[]
	\centerline{\includegraphics[width=0.6\linewidth]{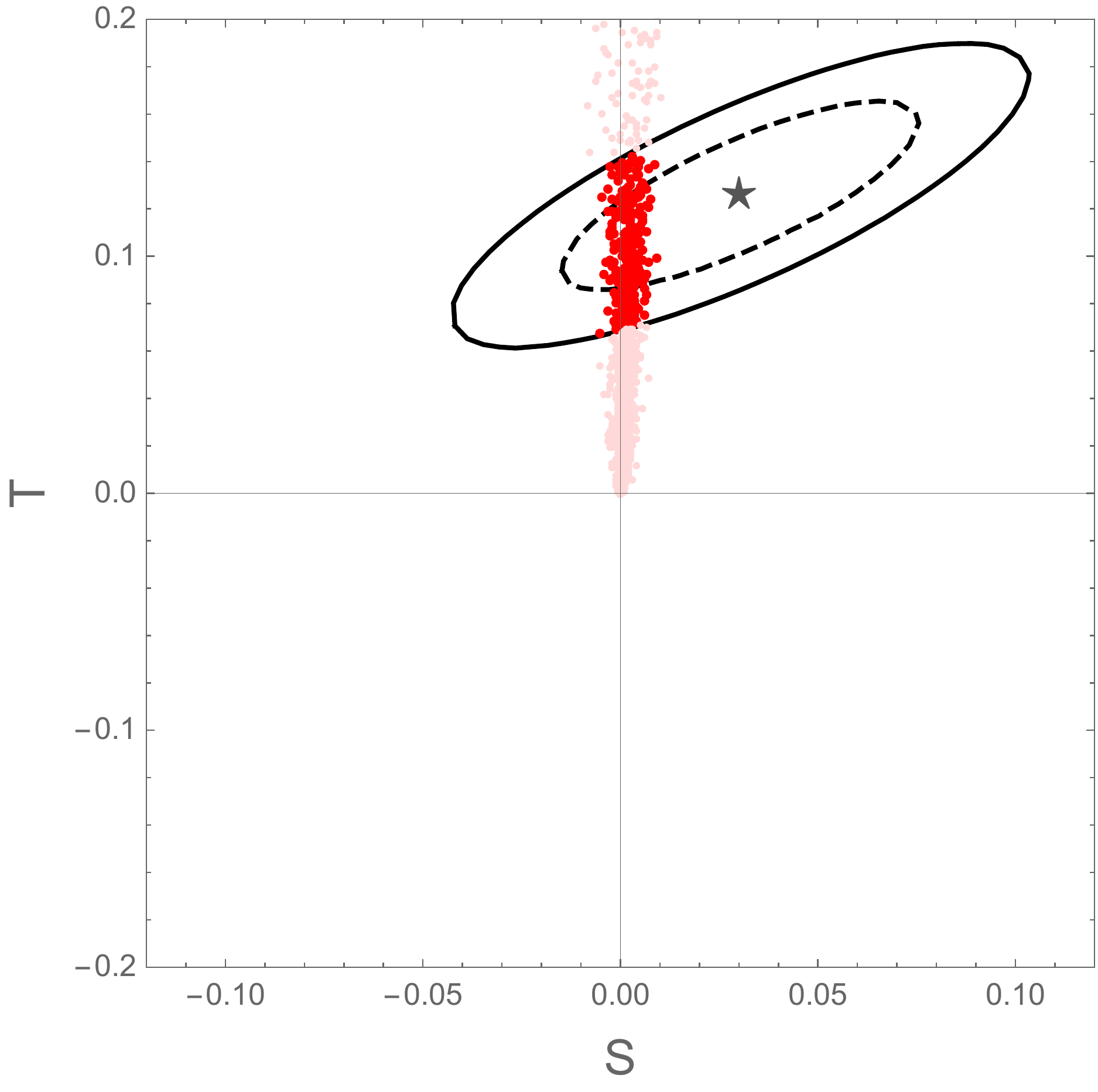}}
	\hspace{0.5cm}
	\caption{ Scatter plot in $(S,T)$-plane. The star mark represents the best fit point, and the dashed (solid) ellipse is
68\% (95\%) contour of fits to the oblique parameters $S$ and $T$ including the CDF result.
The light red and the red points are the ones in Figs.~\ref{fig:trident} accommodating the $\Delta a_\mu$ at the 1$\sigma$ level.
The red points also satisfy the fit at the 95\% CL., which we consider CDF $M_W$-favoured ones. }
\label{fig:ST}
\end{figure}

Fig.~\ref{fig:ST} shows the predictions of $S,T$-parameters for the data points in Fig.~\ref{fig:trident}. The star mark represents the best fit point, and the dashed (solid) ellipse is
68\% (95\%) contour of fits to the oblique parameters $S$ and $T$ including the CDF result~\cite{Strumia:2022qkt,Bagnaschi:2022whn}.
The light red and the red points are the ones in Fig.~\ref{fig:trident} accommodating the $\Delta a_\mu$ at the 1$\sigma$ level.
We can see that the red points also satisfy the fit at the 95\% CL., which we consider CDF $M_W$-favoured ones.
Fig.~\ref{fig:ST} shows that the contribution of the VLLs to the $T$ parameter can be sizable enough to explain the CDF $M_W$ measurement. But the $S$ parameter is
barely affected by the presence of the VLLs.

\begin{figure}[]
	\subfloat[]{\includegraphics[width=0.32\linewidth]{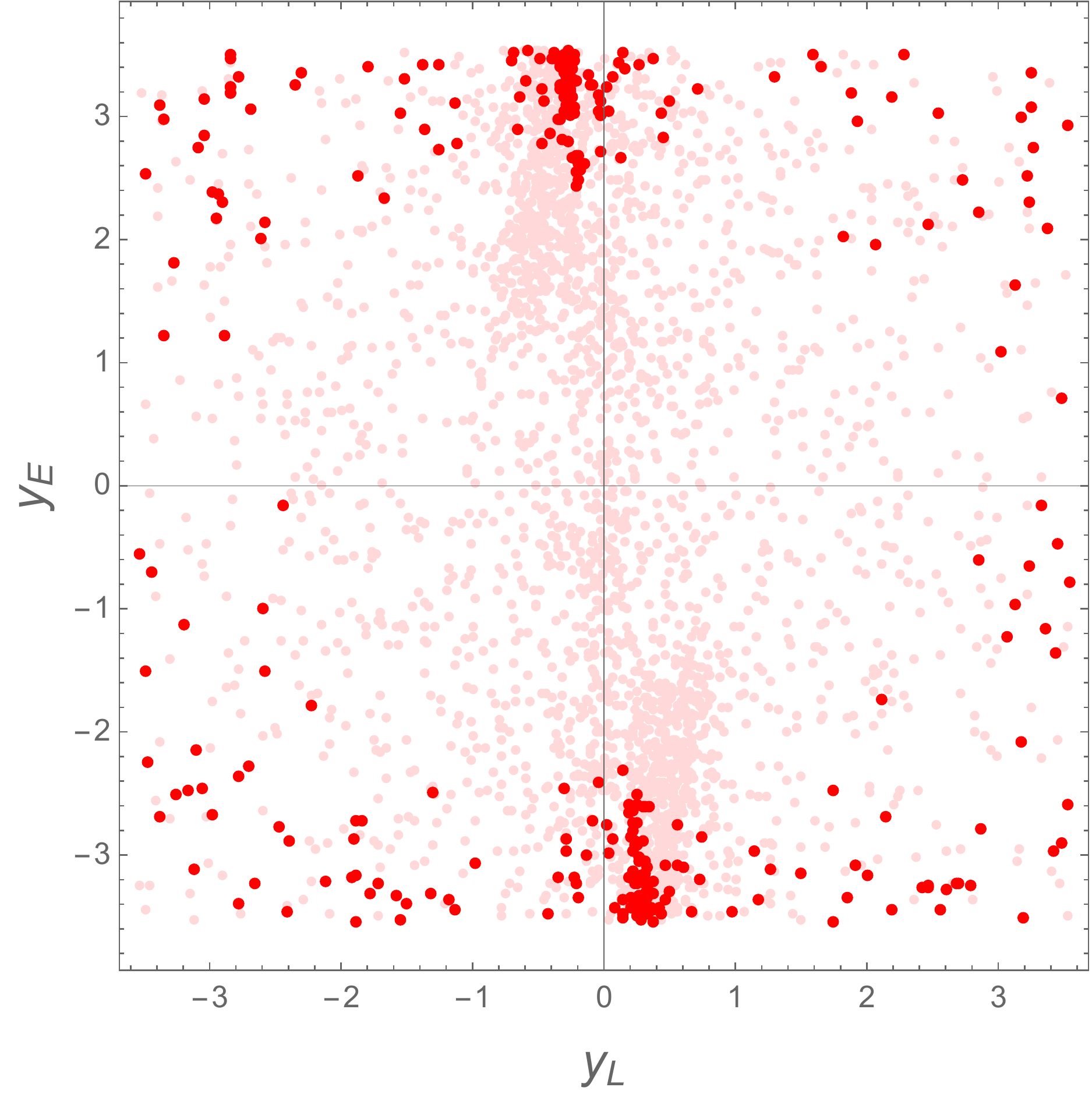}}
	\subfloat[]{\includegraphics[width=0.32\linewidth]{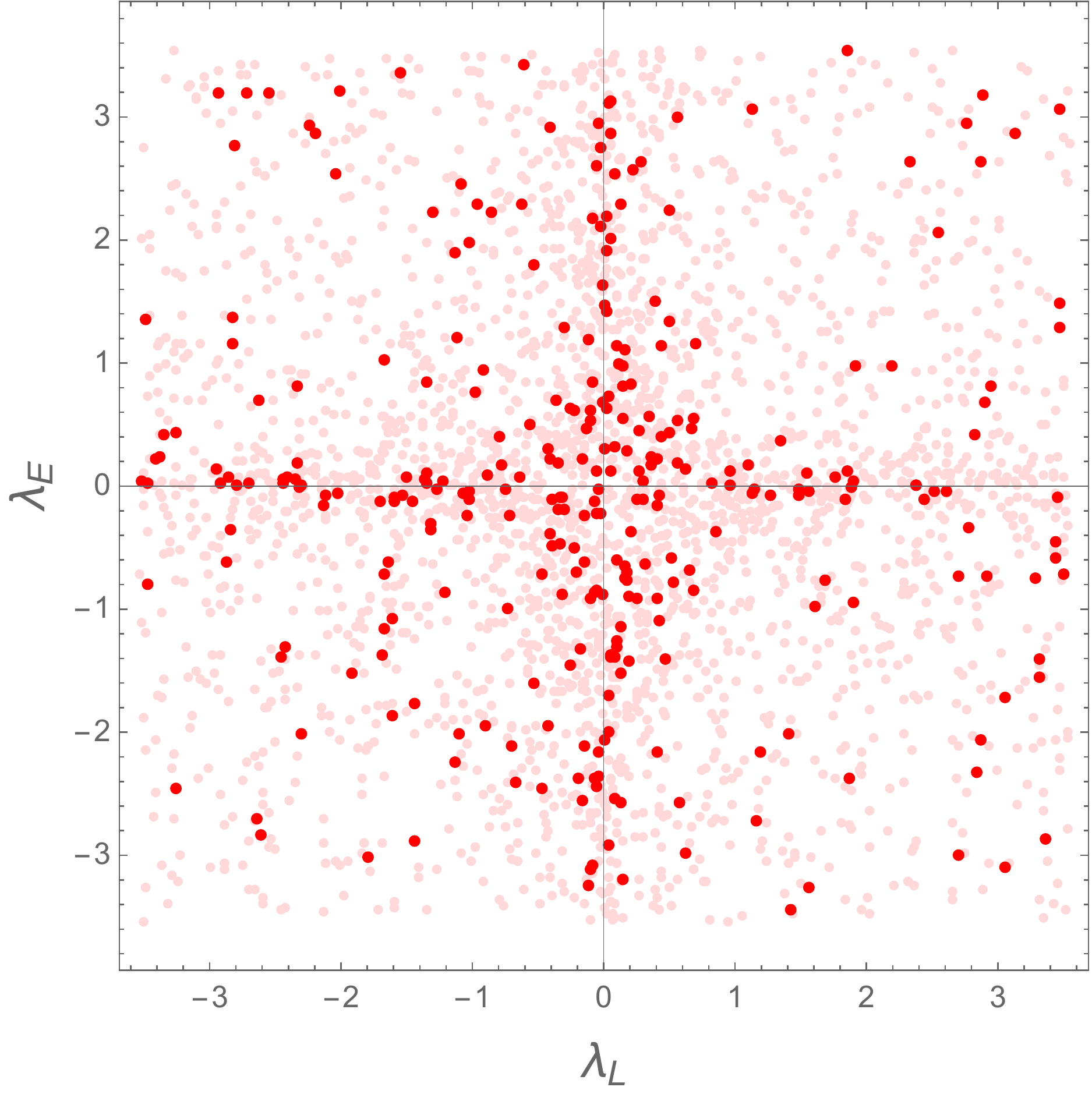}}
	\subfloat[]{\includegraphics[width=0.335\linewidth]{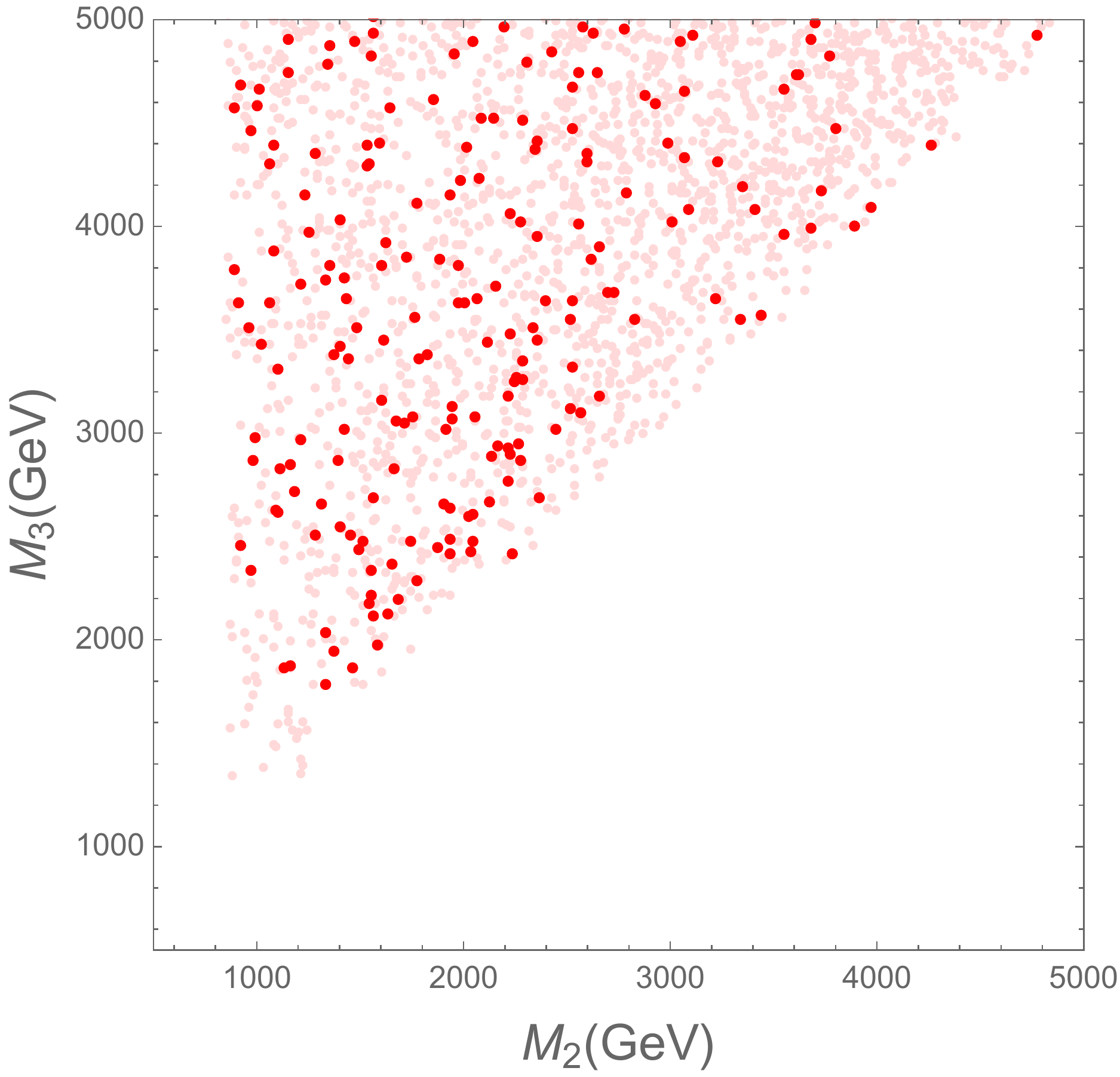}}	
	\hspace{0.5cm}
	\caption{Scatter plot in (a) $(y_L, y_E)$- (b)$(\lambda_L,\lambda_E)$- (c)$(M_2,M_3)$-plane. All the points can explain $\Delta a_\mu$ at 1$\sigma$-level. 
	The red points favour the CDF $M_W$ while the light red points do not. }
\label{fig:y_LE}
\end{figure}

Fig.~\ref{fig:y_LE}(a) shows scatter plot in (a) $(y_L, y_E)$- (b)$(\lambda_L,\lambda_E)$- (c)$(M_2,M_3)$-plane. We use the same data points with the ones in Fig.~\ref{fig:trident}.
The red points favour the CDF $M_W$ while the light red points do not. We can see that the $\Delta a_\mu$ can be accommodated in a wide range of $y_{L(E)}$ while
the CDF $M_W$ favoured points require rather large $y_E$  values although they are still in the perturbative regime. Fig.~\ref{fig:y_LE}(b) shows that neither $\Delta a_\mu$ or CDF $M_W$ does not need large values of $\lambda_{L(E)}$. In Fig.~\ref{fig:y_LE}(c) we can see that although relatively light VLLs are slightly favoured for the CDF $M_W$, even VLLs as heavy as a few TeV can still account for the $\Delta a_\mu$.

\section{Conclusions}
The anomalies in the measurements of the anomalous magnetic moment of muon and the $W$-boson mass, if confirmed by future experiments, suggest the existence of new physics beyond the standard model of particle physics. We considered a possible solution to both of these problems.
We extended the gauge group of  the standard  model to include $U(1)_{L_\mu-L_\tau}$.
We also added a vector-like lepton doublet $L$ and a vector-like lepton singlet $E$ with hypercharge, $-1/2$ and $-1$, respectively.
The extra gauge group is broken by a scalar singlet field $\Phi$. All the new particles are charged under $U(1)_{L_\mu-L_\tau}$. 

We showed that the stringent constraint on the solution to the muon $(g-2)$ in the minimal $U(1)_{L_\mu-L_\tau}$ is significantly lifted.
We also found that the excess of the $W$-boson mass can be explained in the perturbative regime.

\acknowledgments
This work was supported in part by the National Research Foundation of Korea(NRF) grant funded by
     the Korea government(MSIT), Grant No. NRF-2018R1A2A3075605.

\appendix
\section{The treatment of the $\mu$-mass}
\label{sec:mu_mass}
{
Due to the sizable mixings between the muon and the VLLs, $m_{\mu}$ in (\ref{eq:mass_matrix}) can be sizably different from the physical mass $m_{\mu'}$.
Since the muon mass is precisely measured in the experiments, $m'_\mu \approx 0.10565837$ GeV, we use this value as an input parameter. Then the bare muon mass parameter
$m_\mu$ in (\ref{eq:mass_matrix}) can be calculated by solving the characteristic equation,
\begin{align}
\det ( {\cal M}^T {\cal M} -m_{\mu'}^2 \mathbb{1}) =0,
\label{eq:characteristic}
\end{align}
where $\mathbb{1}$ is the $3\times 3$ unit matrix. The (\ref{eq:characteristic}) is a quadratic equation for $m_\mu$, which can be easily solved.
We take smaller positive solution of $m_\mu$.
Then the diagonalization of (\ref{eq:mass_matrix}) by (\ref{eq:M_diag}) guarantees that the lightest eigenvalue corresponds to the experimentally measured muon mass.
}

\section{The $Z(Z')-\mu-E_j$ couplings and the loop-function}
\label{app:Candf}

The $Z^{(\prime)}$-boson interactions with $E_i$ $(i=1,\cdots 3)$ are given by
\begin{align}
{\cal L} \supset - Z^{(\prime)}_\mu \;\overline{E}_i \gamma^\mu \bigg[ C^L_{ij}(Z^{(\prime)}) P_L + C^R_{ij}(Z^{(\prime)}) P_R\bigg] E_j,
\end{align}
where we note $E_1 \equiv \mu'$ and
\begin{align}
C^L_{ij}(Z') &= {g_X} \Bigg[ 2\delta_{ij} -  {V_L}_{1i} {V_L}_{1j}\Bigg], \nl
C^R_{ij}(Z') &= {g_X} \Bigg[ 2\delta_{ij} -  {V_R}_{1i} {V_R}_{1j}\Bigg], \nl
C^L_{ij}(Z) &= { e \over c_W s_W}\Bigg[\left(-{1 \over 2} + s_W^2\right) \delta_{ij} + {1 \over 2} {V_L}_{3i} {V_L}_{3j}\Bigg], \nl
C^R_{ij}(Z) &= { e \over c_W s_W}\Bigg[s_W^2 \delta_{ij} - {1 \over 2} {V_R}_{2i} {V_R}_{2j}\Bigg].
\label{eq:V-m-m}
\end{align}
The loop-function is~\cite{Leveille:1977rc}
\begin{align}
&f(x_,M,{m_{X}}) \nl
&= x (1-x) \left(x + {2 M \over m_\mu} -2\right)-{x^2 \over 2 m_{X}^2}  (M-m_\mu)^2 \left(x-1-{M \over m_\mu}\right).
\end{align}

\section{The vacuum polarization amplitudes}
\label{sec:VP}

The vacuum polarization amplitudes are defined as
\begin{align}
\wt{\Pi}_{LL}(m_1,m_2,0) =&  \frac{m_1^2 + m_2^2}{4} \left(\text{div} + \ln {\mu^2 \over m_1 m_2} \right) +
\frac{m_1^2 + m_2^2}{8} + \frac{m_1^4+m_2^4}{8(m_1^2-m_2^2)} \ln{m_2^2  \over m_1^2}, \nl
\wt{\Pi}_{LR}(m_1,m_2,0) =&  -\frac{m_1 m_2}{2}\left(\text{div}+\ln {\mu^2 \over m_1 m_2} + 1 +\frac{m_1^2+m_2^2}{2(m_1^2-m_2^2)} \ln{m_2^2 \over m_1^2}\right),\nl
\wt{\Pi}'_{LL}(m_1,m_2,0)=& -{1 \over 6}\left(\text{div}+\ln{\mu^2 \over m_1 m_2}\right)
-\frac{m_1^4-8 m_1^2 m_2^2+m_2^4}{18(m_1^2-m_2^2)^2} \nl
& \quad -\frac{(m_1^2+m_2^2)(m_1^4-4m_1^2 m_2^2+m_2^4)}{12(m_1^2-m_2^2)^3}\ln{m_2^2 \over m_1^2},\nl
\wt{\Pi}'_{LR}(m_1,m_2,0) =&  -\frac{m_1 m_2}{2}\left(\frac{m_1^2+m_2^2}{2(m_1^2-m_2^2)^2}+\frac{m_1^2 m_2^2}{(m_1^2-m_2^2)^3} \ln{m_2^2 \over m_1^2}\right),\nl
\wt{\Pi}_{VV}(m_1,m_2,0)=&2 \left(\wt{\Pi}_{LL}(m_1,m_2,0) + \wt{\Pi}_{LR}(m_1,m_2,0)\right), \nl
\wt{\Pi}'_{VV}(m_1,m_2,0)=&2 \left(\wt{\Pi}'_{LL}(m_1,m_2,0) + \wt{\Pi}'_{LR}(m_1,m_2,0)\right), 
\end{align}
where $\text{div}=1/\epsilon+\log 4\pi -\gamma_E$.

 \bibliographystyle{JHEP}
 \bibliography{biblio.bib}

\end{document}